 \documentclass[final,3p,times,twocolumn]{elsarticle}

\usepackage{graphics}
\usepackage{graphicx}
\usepackage[utf8]{inputenc}
\usepackage{indentfirst}
\usepackage{hyperref}
\hypersetup{colorlinks, linkcolor=black, citecolor=black, urlcolor=black}
\usepackage{breakurl}

\journal{arXiv}

\begin{document}
\begin{frontmatter}



\title{Fully-tunable femtosecond laser source in the ultraviolet spectral range}


\author[UNG,Elettra,CEA]{B. Mahieu}\ead{benoit.mahieu@cea.fr}
\author[LUXOR]{S. Coraggia}
\author[Elettra]{C. Callegari}
\author[Elettra,Rome]{M. Coreno}
\author[UNG,Elettra]{G. De Ninno}
\author[Milano]{M. Devetta}
\author[LUXOR]{F. Frassetto}
\author[CEA]{D. Garzella}
\author[Milano]{M. Negro}
\author[Elettra]{C. Spezzani}
\author[Milano]{C. Vozzi}
\author[Milano]{S. Stagira}
\author[LUXOR]{L. Poletto}

\address[UNG]{University of Nova Gorica, Vipavska 11c, 5270 Ajdov\v{s}\v{c}ina, Slovenia}
\address[LUXOR]{Institute for Photonics and Nanotechnologies, CNR-IFN and Department of Information Engineering, University of Padova, 35131 Padova, Italy}
\address[Milano]{Institute for Photonics and Nanotechnologies, CNR-IFN and Department of Physics,
Politecnico di Milano, 20133 Milan, Italy}
\address[Elettra]{Sincrotrone Trieste Elettra, S.S.14 - km 163.5 in AREA Science Park, 34149 Basovizza, Italy}
\address[Rome]{Institute of Inorganic Methodologies and Plasmas, CNR-IMIP, Monterotondo, 00016 Rome, Italy}
\address[CEA]{Service des Photons Atomes et Mol\'{e}cules, Commissariat \`{a}  l'Energie Atomique, Centre d'Etudes de Saclay,
B\^{a}timent 522, 91191 Gif-sur-Yvette, France}


\begin{abstract}
  We demonstrate experimentally the full tunability of a coherent
  femtosecond source in the whole ultraviolet spectral region. The
  experiment relies on the technique of high-order harmonic generation
  driven by a near-infrared parametric laser source in krypton gas. By
  tuning the drive wavelength in the range between 1100 to 1900~nm, we
  generated intense harmonics from near to extreme ultraviolet. A
  number of photons per shot of the order of 10$^7$ has been measured
  for the first harmonic orders. Many novel scientific prospects are
  expected to benefit from the use of such a table-top tunable
  source.
\end{abstract}

\end{frontmatter}
\section{Introduction}
\label{intro}
Generating tunable sub-picosecond radiation at wavelengths shorter
than 250~nm is of great interest to many applications in physics,
chemistry and biology \cite{Haarmalert2009,Wernet2011,Hanley2009}, both in
gas-phase and condensed matter. In fact, below such a wavelength one
overcomes the work function of most solids and clusters of metals and
metalloids, making it possible to eject electrons from the target
sample via optical excitation. The use of short pulses allows one to obtain
important information on the dynamics of fast processes occurring in
systems such as proteins, enzymes and nucleic acids
\cite{Edwards2005}. Such a dynamics is often excited in a given small
range of wavelengths (typically few nanometers). Hence, ``fine''
tunability is an important asset of the employed radiation source. A
more extended tunability (tens of nanometers) is of course also highly
desirable, since it allows the study of a large variety of samples,
with the same radiation source.

Wavelengths below 150~nm are not accessible to conventional lasers,
due to the lack of high-reflectivity broadband mirrors and to the
decreasing efficiency of harmonic conversion in nonlinear crystals
within this spectral range. Synchrotron radiation is not a viable
option either, mainly because pulse duration is generally limited to several
picoseconds. It should be noted that slicing schemes allow the
production of sub-picosecond pulses \cite{Schoenlein2000}, at the cost of a greatly reduced number of photons. Apart from free-electron lasers
\cite{FEL,Allaria2010}, the most effective way to obtain femtosecond coherent pulses
in the ultraviolet and soft X-ray spectral region relies on the
technique of high-order harmonic generation (HHG) in rare gases
\cite{Salieres99,Jaegle,Brabec}. The latter is a highly nonlinear
process based on the interaction of a noble gas with a visible or
infrared (IR) drive laser beam spatially focused so as to reach an intensity
of the order of 10$^{14}$~W/cm$^2$. The outcome of such an
interaction is the generation of a comb of odd harmonics of the laser
wavelength. For the first few harmonic orders (third, fifth,
seventh), the signal intensity falls with the harmonic order and the
process is well explained in the frame of classical nonlinear optics
in centro-symmetric media \cite{New67,Shen}. Higher-order harmonics
instead have almost the same intensity, thus forming a characteristic
plateau. The photon energy at which the plateau's cut-off is located
is proportional to the square of the fundamental wavelength. The HHG
process can be explained by means of the quantum model reported in
\cite{Lewenstein94}. However, the main features of the harmonic
emission can be accounted for using a semiclassical model
\cite{Corkum93,Schafer93}.

Generally, the drive beam for HHG has a fixed fundamental
wavelength, most often around 800~nm for a standard Ti:Sapphire
(Ti:Sa) source. Thus, harmonic wavelengths are also
fixed. Because harmonic peaks are far away from one another,
especially for low harmonic orders, lack of tunability is a serious
drawback for the previously cited scientific investigations: it
narrows accessible optical excitations down to few photon
energies. Driving HHG with a mix of the fundamental wavelength and its
second harmonic is a simple technique to produce both even and odd
harmonics of the fundamental wavelength
\cite{Lambert2009}. However, it only partially fills the gap
between spectral lines, thereby not reaching a full tunability. Full
tunability has been demonstrated for high harmonic orders,
corresponding to extreme-ultraviolet (EUV) wavelengths, with
techniques such as harmonic blueshifting depending on the generation
geometry \cite{Altucci99}, control of the chirp of the drive laser
\cite{Kim2003,Martin2004} or, in the picosecond range, tuning of a non-compressed
narrow-band Ti:Sa drive source over its spectral range of
amplification \cite{Brandi2003}.
 
A more effective and straightforward solution to generate
fully-tunable harmonics is to rely on a widely tunable drive
source. It has been demonstrated in \cite{Bellini2000} for harmonics
around 150~nm. In this paper, we extend this study to the whole
ultraviolet spectral range. The novelty of our work stems from the
unique qualities of the source that we used to drive HHG. This driving source is characterized by a large
wavelength tunability from 1100 to 1900~nm, a mJ-level pulse energy and short pulse duration
of the order of 20~fs. In these conditions the generation of few-femtosecond harmonic radiation is ensured.
Our experiment and its implementation are described in the
next section. Experimental results are then presented and discussed.

\section{Experimental setup}
\label{description}

\begin{figure*}
\centering
\resizebox{0.92\textwidth}{!}{%
  \includegraphics{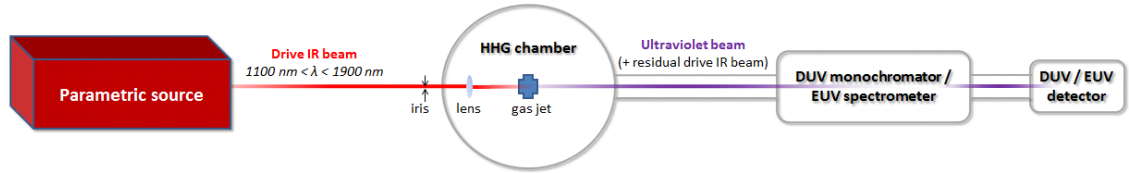}
}
\caption{Layout of the experiment. The parametric source, whose
  comprehensive description can be found in \cite{Vozzi2007}, provides
  a tunable near-IR driving beam. It is focused with a lens of 15~cm
  focal length onto a krypton gas jet at a backing pressure of
  3.5~bars. In order to attain the maximum harmonic yield for a
  given laser focal spot, the position of the jet with respect to the
  laser focus is optimized by carefully adjusting a manual x-y-z
  translation stage with a positioning precision of
  $\pm$0.01~mm. Moreover, an iris placed on the path of the driving
  beam allows to adjust size and intensity at the focus in order to
  optimize the harmonic generation efficiency. The gas is injected in
  the HHG chamber through an electromagnetic valve, synchronized with
  the laser beam, with a nozzle diameter of 0.5~mm operating at an opening
  time of 400 $\mu$s. The spectrum of the harmonic beam is acquired by
  two different detection systems: a monochromator in the DUV and a
  spectrometer in the EUV, with suitable
  detectors. Detection systems in both DUV and EUV regions have been
  calibrated so as to allow measurement of the harmonic absolute
  photon flux.}
\label{setup}
\end{figure*}

The layout of the experiment is shown in Fig.~\ref{setup}. The
parametric source is based on a Ti:Sapphire laser facility providing
intense short pulses (tens of mJ energy; 60~fs duration), centered
at a wavelength of 790~nm, with a repetition rate of 10~Hz. The output
beam stems from difference frequency generation (DFG) of
spectrally broadened pulses. The generated pulses are then amplified
in a two-stage optical parametric amplifier (OPA), leading
to the production of $\sim$20-fs pulses with an energy up to 1.2~mJ,
tunable from 1100 to 1900~nm.  Tunability is achieved by rotating
the crystals in the OPAs, thereby changing the phase-matching
conditions.

The generation of the high-order harmonics of the near-IR driving
pulses is achieved by focusing the laser beam on a jet of krypton gas,
which ensures a better harmonic conversion efficiency than lighter
gases such as argon, at the price of a lower cutoff frequency.

To cover the deep-ultraviolet (DUV) and EUV
spectral regions two different spectrometers have been used. Harmonic
emission in the DUV was analyzed through a normal-incidence
Czerny-Turner scanning monochromator\footnote{McPherson model 218}
equipped with a 2400-gr/mm AlMgF$_2$-coated grating. The
monochromator selects a single harmonic or a spectral portion
thereof. The photon flux at the exit slit of the monochromator is
detected by a photomultiplier tube\footnote{Hamamatsu model R1414}
with a tetraphenyl butadiene (TPB) phosphor photocathode to enhance
the detection efficiency. Owing to the limited spectral range
accessible to the grating, that has significant transmission for
wavelengths above $\sim$130~nm, only the harmonics ranging from the
third to the eleventh order of the fundamental wavelength could be
detected. The harmonic spectra at high resolution were obtained by
scanning the grating, with a 300-$\mu$m slit aperture, giving a
bandwidth of 0.4~nm.

The global response of the instrument (i.e., monochromator plus
detector) has been absolutely calibrated using the facilities
available at CNR-IFN and described in details in \cite{Poletto99}, in
order to measure the DUV photon flux generated in the interaction
region at the different harmonics. This was performed by tuning the
monochromator to one of the harmonics and opening completely its
slits. In such a way, the beam enters the monochromator without
being clipped at the entrance slit and is diffracted by the
grating. The harmonic of interest then exits the monochromator without
being clipped at the output slit, and is detected by the
photomultiplier. We verified that even with the slits completely open
the different harmonics were clearly separated at the output.

The signal in the EUV was analyzed through a grazing-incidence
flat-field spectrometer equipped with a 1200-gr/mm gold-coated grating and
tuned in the 80--35~nm spectral region. The spectrum is acquired by a
40-mm-diameter microchannel plate intensifier with MgF$_2$
photocathode and phosphor screen optically coupled with a low-noise
CCD camera. Also in this case, the global response of the
instrument (i.e., grating and detector) has been absolutely calibrated,
as described in detail in \cite{Poletto2001,Poletto2004}. Since the
spectrometer works without an entrance slit, having the harmonics
generation point as its input source, all the generated EUV photons
enter the instrument and are diffracted onto the detector.

\section{Results and discussions}
\label{results}

\subsection*{Deep-ultraviolet region}

\begin{figure}
\centering
\resizebox{0.5\textwidth}{!}{%
  \includegraphics[trim = 0 0 0 3cm, clip]{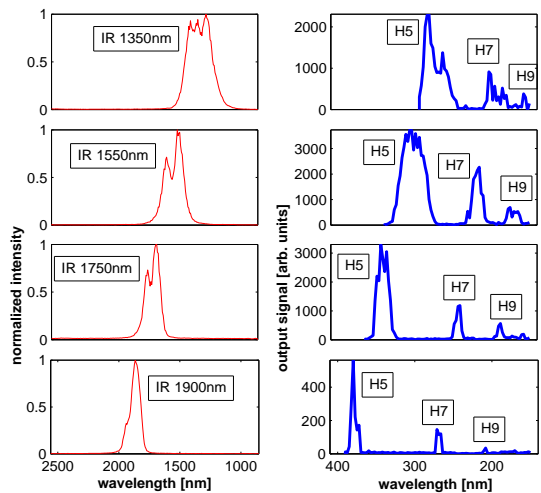}
}
\caption{IR spectra (left side) and corresponding harmonic spectra
  (right side) in the 400--150~nm spectral region. H5, H7 and H9
  stand respectively for the fifth, seventh and ninth harmonics of
  the driving IR beam.}
\label{spectra_DUV} 
\end{figure}

Figure~\ref{spectra_DUV} shows the spectral characterization from four
sets of measurements, corresponding to four different
wavelengths of the driving pulse: 1350, 1550, 1750, and 1900~nm. It is important to
stress that the stability of the beam provided by the parametric
source ensures a very good reproducibility of the measurements.

Below 150~nm, the efficiency of the DUV monochromator is dramatically
low. Thereby, the analysis of harmonic spectra has been done only down
to 150~nm. Since the third harmonics of the considered driving IR
wavelengths are generally located in the visible, i.e., out of the
monochromator range, harmonic orders from fifth to ninth have been
analyzed. As expected at these relatively low orders, the signal
quickly decreases with increasing harmonic order. Indeed, the
intensity of harmonics before the plateau region is related to the
probability of multiphoton ionization of the gas atoms
\cite{Tong2001}. Like the driving beam, harmonics have a large
bandwidth (a few tens of nanometers), intrinsic to an ultrashort pulse
source.

\begin{figure}
\centering
\resizebox{0.5\textwidth}{!}{%
  \includegraphics{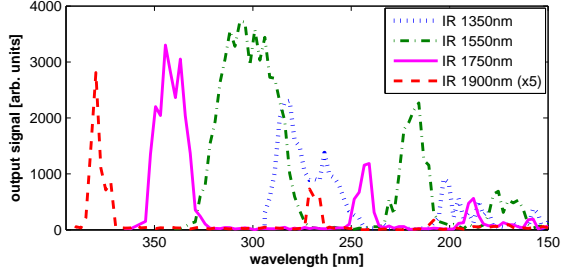}
}
\caption{Overlap of the harmonic spectra for four drive IR wavelengths
  (1350, 1550, 1750 and 1900~nm). The harmonic
  spectrum resulting from the drive wavelength of 1900~nm has been
  vertically magnified ($\times 5)$. }
\label{overlap_DUV} 
\end{figure}

The overlap of the harmonic spectra shows a full tunability of the
source in the DUV spectral region (Fig.~\ref{overlap_DUV}). The range
between 400 and 350~nm corresponds to either the fifth harmonic of a
1750--2000~nm fundamental beam or the third harmonic of a 1050--1200~nm
fundamental beam. These wavelengths are the boundaries of the
accessible spectral range of the used parametric source, so that in
these regions the IR spectrum is less stable and moreover the beam
energy is lower than in the 1350--1550~nm ``peak region''. Hence
harmonics in the 400--350~nm region are also less intense. The third
harmonic of a 1050--1200~nm fundamental beam can be generated with
better conversion efficiency in the frame of classical nonlinear
optics in crystals \cite{Craxton81}.

\begin{figure}
\centering
\resizebox{0.5\textwidth}{!}{%
  \includegraphics{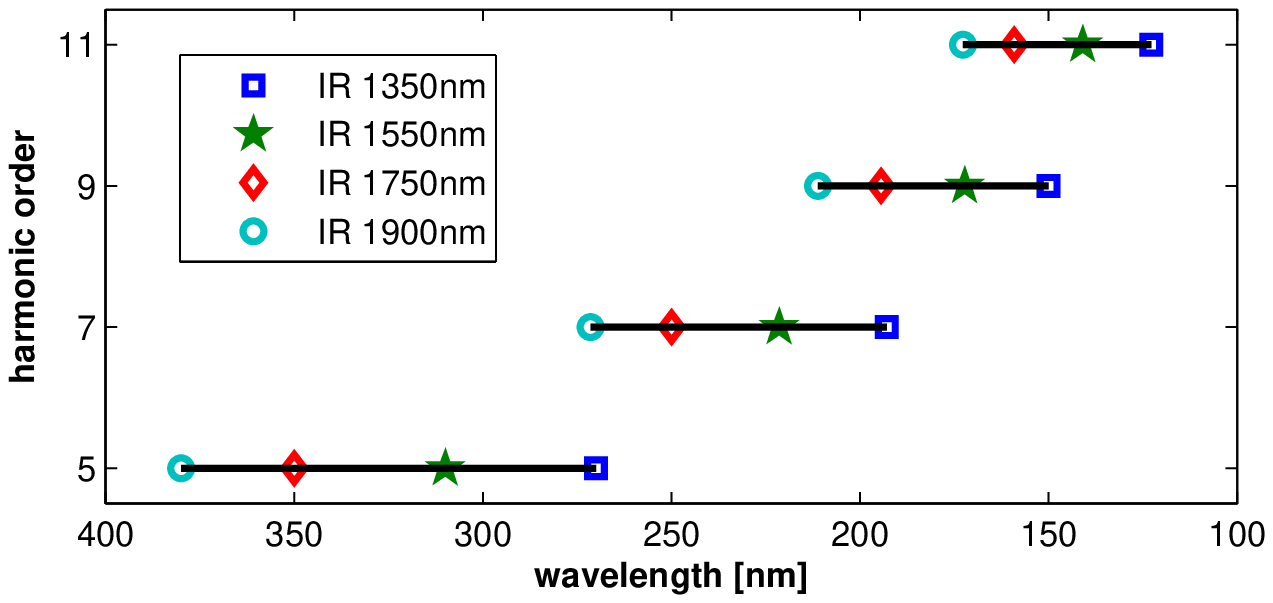}
}
\caption{Tunability in the DUV. The lines represent the wavelength
  ranges that are covered by harmonic orders from fifth to evelenth,
  generated by drive wavelengths ranging from 1350~nm to 1900~nm.}
\label{tunability_DUV} 
\end{figure}

Figure~\ref{tunability_DUV} clearly shows that when the driving
wavelength ranges from 1350 to 1900~nm, as in these measurements,
harmonic orders from fifth to eleventh completely cover the DUV spectral region. Furthermore, the third harmonic, not shown
in Fig.~\ref{tunability_DUV}, also allows tunability in the visible
region. Obviously this overlap and thereby the tunability in the
ultraviolet range improve at shorter wavelengths, where narrower IR
tunability is thus sufficient.

The photon flux of the harmonics has been measured by fully opening
the slits of the monochromator in order to get all the signal on the
photomultiplier. The results are summarized in Table
\ref{efficiencyVUV}.
\begin{table}
\centering
\caption{Absolute number of photons provided in the DUV spectral range for
  four different drive wavelengths. Left column: central
  wavelength of the drive beam; center column: wavelengths corresponding 
  to the peak signal of the indicated
  harmonics; rigth column: measured photons/shot for each harmonic.}

\label{efficiencyVUV}
\begin{tabular}{c|c|c}
  drive IR & harmonic & photons/shot \\
  {[nm]} & [nm (order)] &  \\
  \hline
  1160 & 414 $(3^{\mathrm {rd}})$ & $2.0 \times 10^8$\\
  & 248 $(5^{\mathrm {th}})$ & $6.1 \times 10^7$\\
  & 177 $(7^{\mathrm {th}})$ & $3.6 \times 10^7$\\
  \hline
  1350 & 270 $(5^{\mathrm {th}})$ & $4.2 \times 10^7$\\
  & 196 $(7^{\mathrm {th}})$ & $1.8 \times 10^7$\\
  & 153 $(9^{\mathrm {th}})$ & $1.3 \times 10^7$\\
  \hline
  1450 & 285 $(5^{\mathrm {th}})$ & $6.1 \times 10^7$\\
  & 204 $(7^{\mathrm {th}})$ & $1.6 \times 10^7$\\
  & 161 $(9^{\mathrm {th}})$ & $8.2 \times 10^6$\\  
  \hline
  1800 & 367 $(5^{\mathrm {th}})$ & $1.3 \times 10^7$\\
  & 262 $(7^{\mathrm {th}})$ & $7.0 \times 10^6$\\
  & 206 $(9^{\mathrm {th}})$ & $3.4 \times 10^6$\\    
\end{tabular}
\end{table} 
Around 10$^7$ photons per shot are generated in the DUV spectral
region, corresponding to a beam energy of the order of 10~pJ. One sees
that the higher the driving wavelength, the smaller the harmonic photon
flux. Regarding the ninth harmonic of the driving laser for 1350, 1450 and 1800~nm
drive wavelengths ($\lambda _{\mathrm{IR}}$), the harmonic conversion
efficiency scales as $\lambda _{\mathrm{IR}}^{-6.34}$. This is in
agreement with recent theoretical studies which show that the harmonic
efficiency in the plateau region scales as $\lambda _{IR}^{-6}$, not
as $\lambda _{\mathrm{IR}}^{-3}$, as previously believed
\cite{Tate2007}. Moreover, in \cite{Shiner2009}, the conversion
efficiency of further plateau harmonics (from 78 to 39~nm) has
recently been measured to be proportional to $\lambda
_{\mathrm{IR}}^{-6 \pm 1.1}$ in krypton. Although increasing the driving
wavelength allows to extend the harmonic plateau
\cite{Popmintchev2008}, there is a penalty in terms of harmonic
efficiency.

\subsection*{Extreme-ultraviolet region}

\begin{figure}
\centering
\resizebox{0.5\textwidth}{!}{%
  \includegraphics[trim = 0 0 0 2cm, clip]{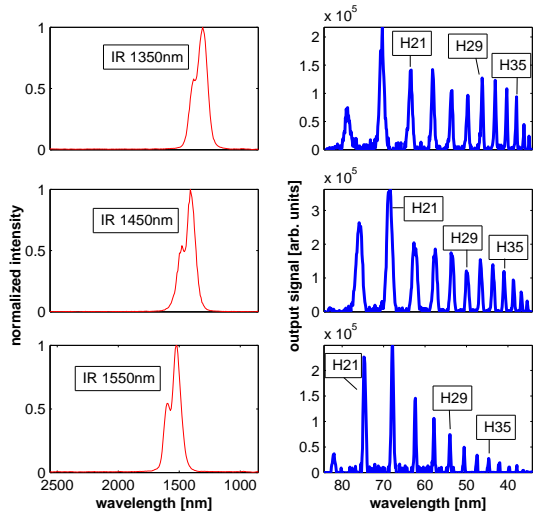}
}
\caption{IR spectra (left side) and corresponding harmonic spectra
  (right side) in the 85--30~nm spectral region. H21, H29 and H35
  stand respectively for the $21^{\mathrm{st}}, 29^{\mathrm{th}},
  35^{\mathrm{th}}$ harmonic order of the drive IR beam.}
\label{spectra_EUV} 
\end{figure}

\begin{figure}
\centering
\resizebox{0.5\textwidth}{!}{%
  \includegraphics{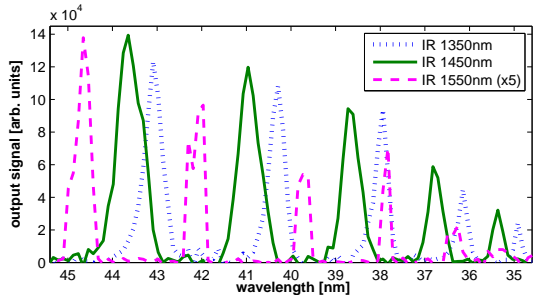}
}
\caption{Overlap of the harmonic spectra in the range 45--35~nm for
  three drive IR wavelengths (1350, 1450 and 1550~nm). The harmonic
  spectrum resulting from the drive wavelength of 1550~nm has been
  vertically magnified ($\times 5)$.}
\label{overlap_EUV} 
\end{figure}

\begin{figure}
\centering
\resizebox{0.5\textwidth}{!}{%
  \includegraphics{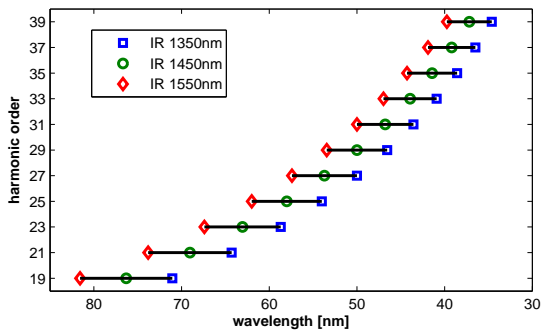}
}
\caption{Tunability in the EUV. The lines represent the wavelength
  ranges that are covered by harmonic orders from number nineteen to
  thirty-nine, generated by drive wavelengths ranging from 1350~nm to
  1550~nm.}
\label{tunability_EUV} 
\end{figure}

The same procedure has been followed
for the measurements performed in the EUV region, using the 
detection system described before. Harmonic spectra are reported in
Fig.~\ref{spectra_EUV} for three different driving wavelengths (1350,
1450, 1550~nm) and their overlap in the 45--35~nm spectral range is
shown in Fig.~\ref{overlap_EUV}. Figure~\ref{tunability_EUV} shows that
by varying the driving wavelength from 1350~nm to 1550~nm one attains the full tunability in the EUV, through relatively high-order
harmonics. An interesting point is that one specific ultraviolet
wavelength can be obtained from multiple drive wavelengths through
different harmonic orders.

The absolute number of photons in the EUV is reported in Table
\ref{efficiencyEUV} for harmonic orders 21, 29 and 35 of 1350, 1450
and 1550~nm driving wavelengths. Such a photon flux corresponds to
an energy per harmonic per shot about two orders of magnitude
smaller than in the DUV. This can be explained by the different nature of the harmonic generation process and of phase matching conditions in the DUV and EUV spectral regions.
\begin{table}
\centering
\caption{Absolute number of photons provided in the EUV spectral range
  for three different drive wavelengths. Left column: central wavelength of the drive beam; center column: wavelengths of
  harmonic orders 21, 29 and 35; right column: measured
  photons/shot for each harmonic.}
\label{efficiencyEUV}
\begin{tabular}{c|c|c}
drive IR & harmonics & photons/shot \\
  {[nm]} & [nm] &  \\
\hline
1350 & 63 & $1.3 \times 10^4$\\
& 46 & $5.8 \times 10^3$\\
& 38 & $2.8 \times 10^3$\\
\hline
1450 & 69 & $4.4 \times 10^4$\\
& 50 & $9.9 \times 10^3$\\
& 41 & $4.7 \times 10^3$\\
\hline
1550 & 75 & $1.4 \times 10^4$\\
& 54 & $3.1 \times 10^3$\\
& 45 & $6.5 \times 10^2$\\
\end{tabular}
\end{table} 

Different  strategies can be pursued to overcome this low photon
flux. As a first possibility, one could design a more powerful
parametric source \cite{Takahashi2008}. A
complementary strategy is the improvement of the HHG process in terms
of tunability and conversion efficiency. In this respect, a promising
technique that could be investigated is mixing the fundamental
wavelength of the parametric source with either its second harmonic or
with a standard powerful Ti:Sa laser source, as demonstrated in \cite{Calegari2009}.

\section{Conclusion}
\label{conclusion}

The full tunability of a femtosecond photon beam produced through HHG
driven by a parametric source has been demonstrated in the whole
ultraviolet spectral range. This source opens the way to novel
scientific experiments. The main drawback comes from the relatively
low harmonic conversion efficiency, resulting from a drive wavelength
longer than in classic HHG setups.  Increasing the harmonic photon
flux would not only extend the range of possible scientific
experiments, but also pave the way for the development of a tunable
ultraviolet/soft X-ray source for seeding single-pass free-electron
lasers, an application for which HHG-based sources have demonstrated to be attractive \cite{Lambert2008}.\\\

\section*{Acknowledgements}
The authors thank Luca Romanzin (Sincrotrone Trieste) and  Andrea Martin (CNR-IOM) for technical assistance in the installation of the spectrometer.

This work has been supported by the CITIUS project
\cite{interreg} of the Italian-Slovenian Crossborder Cooperation Programme.



%
%

\end{document}